\documentclass[conference]{IEEEtran}
\usepackage{amsmath,amssymb,amsfonts}
\usepackage[linesnumbered,ruled,vlined,commentsnumbered]{algorithm2e}
\usepackage{graphicx}
\usepackage{textcomp}
\usepackage{xcolor}
\def\BibTeX{{\rm B\kern-.05em{\sc i\kern-.025em b}\kern-.08em
    T\kern-.1667em\lower.7ex\hbox{E}\kern-.125emX}}

\begin{document}

\title{\Huge Study of Polar Codes Based on Piecewise Gaussian Approximation\\
\thanks{This work was supported by CNPq, CAPES, FAPERJ, and FAPESP.}
}

\author{\IEEEauthorblockN{Robert M. Oliveira and Rodrigo C. de Lamare}
\IEEEauthorblockA{\textit{\rm Centre for Telecommunications Studies (CETUC)} \\
{Pontifical Catholic University of Rio de Janeiro (PUC-Rio),
Rio de Janeiro-RJ, Brasil} \\
Emails: rbtmota@gmail.com and delamare@cetuc.puc-rio.br }
}

\maketitle

\begin{abstract}
In this paper, we investigate the construction of polar codes by Gaussian approximation (GA) and develop an approach based on piecewise Gaussian approximation (PGA). In particular, with the piecewise approach we obtain a function that replaces the original GA function with a more accurate approximation, which results in significant gain in performance. The proposed PGA construction of polar codes is presented in its integral form as well as an alternative approximation that does not rely on the integral form. Simulations results show that the proposed PGA construction outperforms the standard GA for several examples of polar codes and rates. \\
\end{abstract}


\section{Introduction}

Polar codes were originally proposed by Arikan in \cite{Arikan}. Due to their low complexity encoding and decoding characteristics, polar codes have quickly become popular and were selected by the 3GPP Group for the uplink/downlink channel control in the 5G \cite{5G} standard.

The construction of polar codes consists of how to choose the best channels to transmit bits of information \cite{Arikan}, called noiseless channels, or the other way around, find the noisy channels. Several design methods have been proposed in the last decade such as parameter Bhattacharyya and Monte Carlo \cite{Arikan}; density evolution (DE) \cite{Mori1},\cite{Mori2},\cite{Tal1}; Gaussian approximation (GA) of density evolution \cite{Chung},\cite{Trifonov}; and the polarization weight (PW) \cite{RZhang}\cite{Schurch}. In particular, the Gaussian approximation (GA) has become the most adopted construction approach due to its low computational complexity when compared to the other methods.

The GA construction method was originally proposed in \cite{Chung} for LDPC codes and in \cite{Trifonov} we have its first application for polar codes. Alternatively to its integral format and complex integration required, the authors in \cite{Chung} and \cite{Jeongseok} proposed approximate GA (AGA) for reducing the computational complexity. The authors in \cite{Fang}, \cite{Dai} and \cite{Ochiai} proposed improved Gaussian approximations for long  blocks with better performance than AGA. Examples of algorithms for implementing the GA can be found in \cite{Vangala} and \cite{Cheng}.

In this paper, we investigate the construction of polar codes by Gaussian approximation (GA) and develop an approach based on piecewise Gaussian approximation (PGA). In particular, with the piecewise approach method we obtain a function that replaces the original GA function with a more accurate approximation, which results in significant gain in performance. The proposed PGA design of polar codes is presented in its integral form as well as an alternative approximation that does not rely on the integral form. The PGA design can have an impact on the design of rate-compatible polar codes \cite{Oliveira1},\cite{Oliveira2},\cite{Oliveira3},\cite{Oliveira4}; which are techniques that use GA in their construction. Simulations results show that the proposed PGA design outperforms the standard GA for several examples of polar codes and rates.

This article is structured with the following sections. In Section II, we have the basic definitions of PC, the notation used, the method of encoding and decoding and a brief summary of GA construction. In Section III, we present the construction of polar codes by approximate Gaussian approximation. In Section IV, we show the comparative simulations and a Section V we find the conclusions of this work.

\section{Polar Codes}

Given a symmetric binary-input, discrete and memoryless channels (B-DMC) $W:\mathcal{X} \to \mathcal{Y}$, where $\mathcal{X}$  $\in \{0,1\}$ and $\mathcal{Y} \in \mathbb{R}$. We have that $W(y|x)$ is the channel transition probability, with $x \in \mathcal{X}$ and $y \in \mathcal{Y}$. In order to transmit the information bits, the most reliable sub-channels are chosen, represented by $K$. $\mathcal{A}$ is the set of $K$ indices. In turn, $\mathcal{A}^\text{c}$ is its complementary set, containing the indices of the least reliable channels (frozen bit sequence). Polar codes can be completely specified by three parameters, PC$(N,K,\mathcal{A}^\text{c})$, where $N$ is the codeblock length, $K$ the length of the information sequence. Define rate as $R=K/N$.

Denote $W^N: \mathcal{X}$ $^N$ $\to \mathcal{Y}$ $^N$ with
\begin{equation}
  W^N(y_1^N|x_1^N)=\prod_{i=1}^N W(y_i|x_i)
\label{eq:01}
\end{equation}

The mutual information is defined by \cite{Arikan}
\begin{equation}
  I(W) = \sum\limits_{y \in Y}\sum\limits_{x \in X}\frac{1}{2}W(y|x)\log\frac{W(y|x)}{\frac{1}{2}W(y|0)+\frac{1}{2}W(y|1)},
\label{eq:02}
\end{equation}
where the base-2 logarithm $0 \leq I(W) \leq 1$ is employed.

On the $N$ independent channels of $W$ we apply the polarization process \cite{Arikan}, we obtain a set of polarized channels $ W_N ^ {(i)}: \mathcal{X} \to \mathcal{Y} \times \mathcal{X}^{\text{i-1}}$, $i = 1,2,\ldots,N$. As defined in \cite{Arikan}, this channel transition probability is given by
\begin{equation}
  W_N^{(i)}(y_1^N,u_1^{(i-1)}|u_i) = \sum\limits_{u_{i+1}^N \in X^{N-1}}\frac{1}{2^{N-1}}W_N(y_1^N|u_1^N).
\label{eq:03}
\end{equation}

According to \cite{Arikan}, $N \to \infty$, {$I(W_N^{(i)})$} tends to $0$ or $1$.

\begin{figure}[htb]
\begin{center}
\includegraphics[scale=0.9]{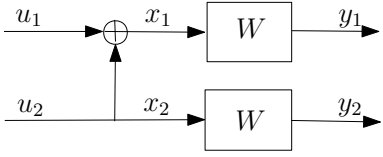}
\caption{The Channel $W_2$}
\end{center}
\label{figura:fig01}
\vspace{-1em}
\end{figure}

In Fig. 1 we show the process of creating the channel $W_2$: recursion step of combining two copies of $W$ independent, that is, $\mathcal{X}^\text{2} \to \mathcal{Y}^\text{2}$ which has the transition probabilities given by
\begin{equation}
  W_2^{(1)}(y_1^2|u_1)=\sum_{u_2}\frac{1}{2}W(y_1|u_1 \oplus u_2)W(y_2|u_2).
\label{eq:04}
\end{equation}
\begin{equation}
  W_2^{(2)}(y_1^2|u_1|u_2)=\frac{1}{2}W(y_1|u_1 \oplus u_2)W(y_2|u_2).
\label{eq:05}
\end{equation}

\subsection{Encoding}

The encoding is given by ${\mathbf x}^N_1 = {\mathbf u}^N_1\textbf{G}_N$, where $\textbf{G}_N$ is the transformation matrix. ${\mathbf u}^N_1 \in \{0.1 \}^N$ is the input block. ${\mathbf x}^N_1 \in \{0.1 \}^N$ is the codeword, where ${\mathbf u}^N_1 = [{\mathbf u}_{\mathcal{A}},{\mathbf u}_{\mathcal{A}^ \text{c}}]$, with ${\mathbf u}_{\mathcal{A}}$ are bits of information and ${\mathbf u}_{\mathcal{A}^\text{c}}$ are frozen bits. We define $\textbf{G}_N = \textbf{B}_N\textbf{F}^{\otimes n}_2 $, where $\otimes$ denotes the Kronecker product, $\textbf{F}_2 = \footnotesize\left[\begin{array}{cc}
1 & 0 \\
1 & 1 \end{array} \right]$ and $\textbf{B}_N$ is the bit-reversal permutation matrix. A simplification without loss of generalization is the omission of $\textbf{B}_N$.

\subsection{Decoding}





Given the received vector ${\mathbf y}^N_1=(y_1,\ldots,y_N)\in
\mathbb{R}^N$, the objective of the decoder is to obtain estimates
of the input of the channel $\hat{\mathbf u}^N_1 \in \{0.1 \}^N$
that is given in vector form as $\hat{\mathbf
u}^N_1=(\hat{u}_1,\ldots,\hat{u}_N)$. The likelihood ratio (LR) of
$u_i$, ${\rm LR}(u_i)=\frac{W^{(i)}_N({\mathbf
y}^N_1,\hat{u}_1^{i-1}|0)}{W^{(i)}_N({\mathbf
y}^N_1,\hat{u}_1^{i-1}|1)}$, using Successive Cancellation (SC)
decoding \cite{Arikan}. Then, the value of the $\hat{u_i}$ is:
\begin{equation}
  \hat{u}_i=\begin{cases}
    h_i({\mathbf y}_1^N,\hat{u}_1^{i-1}), & \text{if $i \in \mathcal{A}$}.\\
    u_i, & \text{if $i \in \mathcal{A}$$^c$},
  \end{cases}
\label{eq:06}
\end{equation}
where $h_i:\mathcal{Y}^\text{N} \times \mathcal{X}$ $^{i-1} \to \mathcal{X}$, $i \in \mathcal{A}$, are decision functions defined as
\begin{equation}
  h_i({\mathbf y}_1^N,\hat{u}_1^{i-1})=\begin{cases}
    0, & \text{if $\frac{W^{(i)}_N({\mathbf y}^N_1,\hat{u}_1^{i-1}|0)}{W^{(i)}_N({\mathbf y}^N_1,\hat{u}_1^{i-1}|1)}\geq 1$}\\
    1, & \text{otherwise}.
  \end{cases}
\label{eq:07}
\end{equation}
for ${\mathbf y}_1^N \in \mathcal{Y}^\text{N}$, $\hat{u}_1$ $^{i-1} \in \mathcal{X}$ $^{i-1}$.

We denote $L(i, j)$ as LR node, $i$ being the line and $j$ being the stage, following mapping of the decoding tree \cite{Arikan}. The values assumed by $L(i,j)$ can be obtained recursively \cite{Arikan} using the equations:
\begin{equation}
  L(i,j+1)= \\
   \begin{cases}
    f(L(i,j),L(i+n/2^{j+1},j)), &\text{($f$ nodes)}\\
    g(L(i-n/2^{j+1},j),L(i,j),\hat{u}_{sum}), &\text{($g$ nodes)}
  \end{cases}
\label{eq:08}
\end{equation}
where $f$ and $g$ functions were defined in \cite{Arikan} as:
\begin{equation}
  f(a,b)=\frac{1+ab}{a+b}
\label{eq:09}
\end{equation}
\begin{equation}
  g(a,b,\hat{u}_{sum})=a^{1-2\hat{u}_{sum}}b
\label{eq:10}
\end{equation}
where $\hat{u}_{sum}$ is the previous decoded bits. The $\Hat{u}_{sum}$ estimated value is given by \eqref{eq:06}. So, the decision $g$ nodes depends on the estimate of $f$ nodes given by \eqref{eq:09}, that is, of previously decoded bits. Message passing decoding that has found numerous applications in wireless communications \cite{bfpeg,rrser,rootldpc,memd,baplnc,dopeg,jidf,spa,mfsic,dfcc,mmimo,mbdf,mbthp,wence,bfidd,vfap,kaids,1bitidd,did,lrcc,aaidd,listmtc,dynovs,rcpd,detmtc,srbars,dynmtc,nupd} can also be considered.


\subsection{Construction by Gaussian Approximation}

The Gaussian approximation (GA) construction method \cite{Chung},\cite{Trifonov} is given by

\begin{equation}
E(L^{(2i-1)}_N) = \phi^{-1}(1-(1-\phi(E(L^{(i)}_{N/2})))^2) \\
\label{eq:11}
\end{equation}
\begin{equation}
E(L^{(2i)}_N) = 2E(L^{(i)}_{N/2}),
\label{eq:12}
\end{equation}
with
\begin{equation}
L^{(0)}_1 = \frac{2}{\sigma^2} 
\label{eq:13}
\end{equation}

The $L_N^{(i)}$ to denote the LLR of the subchannel $W_N^{(i)}$, $\sigma^2$ is the variance and $E[\cdot]$ represents the mean. In practice and for building the algorithm, $E[L_N^{(i)}]=L_N^{(i)}$. The function $\phi(x)$ is defined as:
\begin{equation}
\phi(x) =
\begin{cases}
1-\frac{1}{\sqrt{4\pi x}}\int \limits_{-\infty}^{+\infty} {\rm tanh}(\frac{u}{2})e^{\frac{-(u-x)^2}{4x}} \mathrm{d}u & \ x > 0 \\
1 & \ x = 0 \\
\end{cases}
\label{eq:14}
\end{equation}
Due to the integral form, we call it an exact Gaussian approximation (EGA).

In the polar codes construction we generate the parameter $\mathcal{A}^\text{c}$, depending on $N$, $K$, the type of channel, the target signal-to-noise ratio (SNR), called design-SNR, and the type of decoder. All construction methods covered in this paper are specific to the Additive White Gaussian Noise (AWGN) channel and the SC decoder but can be extended to other types of channels and other type of decoders.

\section{Proposed Construction based on Piecewise Gaussian Approximation}

Observing equation \eqref{eq:14}, we define a new function for analysis:
\begin{equation}
\psi(x,u) = \text{tanh}( \frac{u}{2}) \frac{1}{\sqrt{4 \pi x}} e^{\frac{-(u-x)^2}{4x}}
\label{eq:15}
\end{equation}
we have that the function is the product of a Gaussian function with a hyperbolic tangent function (tanh). Fig. 2 shows the two functions, tanh and Gaussian.

\begin{figure}[htb]
\vspace{-1 em}
\begin{center}
\includegraphics[scale=0.45]{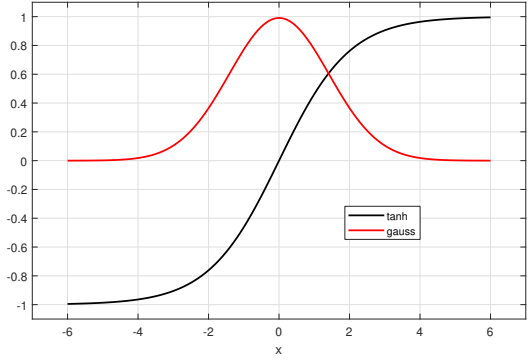}
\vspace{-1 em}
\caption{Example of the tanh function and the Gaussian function.}
\end{center}
\label{fig:01}
\end{figure}

We observe that the tanh function is an odd and zero-centered function. In Fig. 3 we can see the compound function \eqref{eq:15}, which we will call the modified Gaussian and its behavior when we vary the mean, since tanh does not depend on the mean. Each curve is a Gaussian function with an mean ranging from 0 to 15. We can see that the tanh function is dominant over the Gaussian function for $u < 1$. And with the increase in the mean value, the final behavior tends to be a purely Gaussian function. As noted earlier, with an increase in the mean the Gaussian part becomes more relevant. And the tanh function is dominant over the Gaussian one for a small mean and tends to zero.

We propose the PGA, a piecewise approximation for the function $\phi$ in \eqref{eq:14}, which is slightly modified to improve the performance of the GA construction for polar codes. To this end, a new function is proposed to replace the tanh function. For this approximation we will use the exponential function in the following way, $a \cdot e^{(b \cdot x)}$.

\begin{figure}[htb]
\vspace{-1 em}
\begin{center}
\includegraphics[scale=0.45]{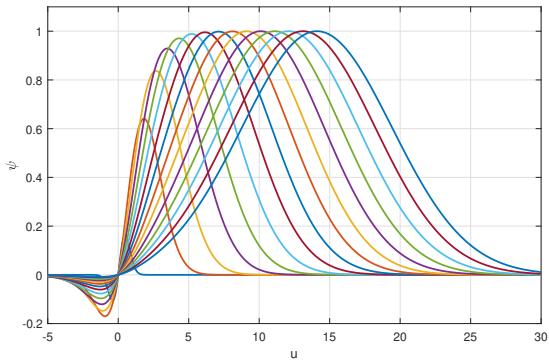}
\vspace{-1 em}
\caption{Function $\psi$ in \eqref{eq:15}.}
\end{center}
\label{fig:01}
\end{figure}
\vspace{-1 em}
The proposed piecewise $\phi_p$ function optimized for PC is
\begin{equation}
\phi_p(x) =
\begin{cases}
1-\frac{1}{\sqrt{4\pi x}}\int \limits_{-\infty}^{+\infty} f(\frac{u}{2})e^{\frac{-(u-x)^2}{4x}} \mathrm{d}u & \ x > 0 \\
1 & \ x = 0
\end{cases}
\label{eq:16}
\end{equation}
with
\begin{equation}
f(x) =
\begin{cases}
 a \cdot e^{(b \cdot x)} + c \cdot e^{(d \cdot x)} & \ x \geq -3.1 \ e \ x \leq 3.1 \\
+1 & \ x > +3.1 \\
-1 & \ x < -3.1. \\
\end{cases}
\end{equation}

The parameters are: $a = 1.9e+07$, $b = 8.4e-09$, $c = -1.8e+07$ and $d = -8.5e-09$; in its integral form. Using the same format and limits proposed by \cite{Jeongseok}, we develop an approximation to the proposed function $\phi_p$ in \eqref{eq:16} given by

\begin{equation}
\phi_p(x) \approx
\begin{cases}
e^{-0.0484x^2-0.3258x} & \ 0 \leq x < 0.867861 \\
e^{-0.4777x^{(0.8512)}+0.1094} & \ 0.867861 \leq x < 10 \\
\sqrt{\frac{\pi}{x}}(1-\frac{1.509}{x})e^{-\frac{x}{3.936}} & \ x \geq 10 \\
\end{cases}
\label{eq:17}
\end{equation}

In Fig. 4 and Fig. 5, we can see the functions in \eqref{eq:11} with the  $\phi$ for GA in \eqref{eq:14} and the $\phi_p$ for PGA in \eqref{eq:16}.

\begin{figure}[htb]
\vspace{-1 em}
\begin{center}
\includegraphics[scale=0.45]{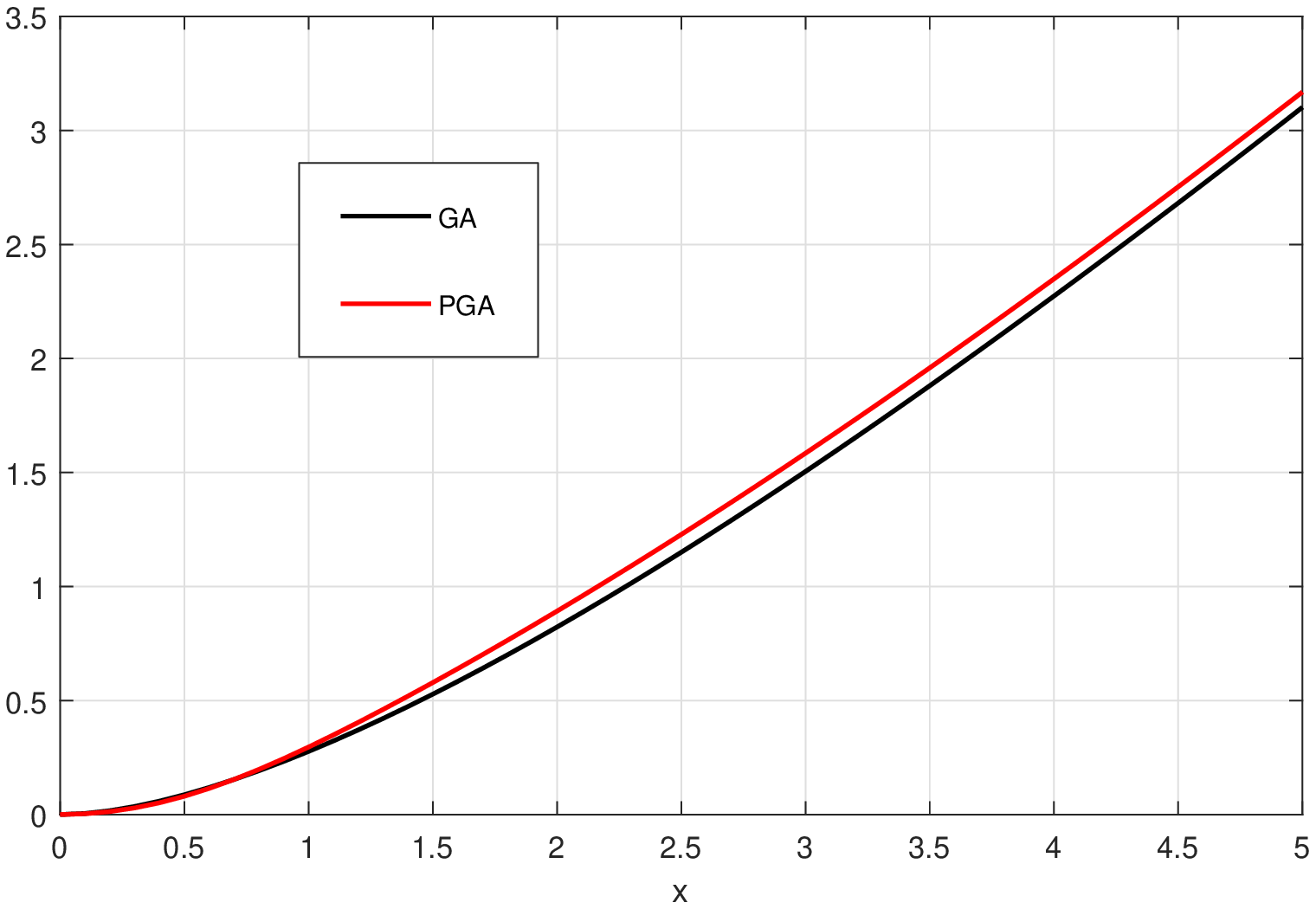}
\vspace{-1 em}
\caption{Comparison of eq. \eqref{eq:11} for GA and PGA with x $\in$ (0,..,5)}
\end{center}
\label{fig:funcao_phi5}
\end{figure}

\begin{figure}[htb]
\vspace{-1 em}
\begin{center}
\includegraphics[scale=0.45]{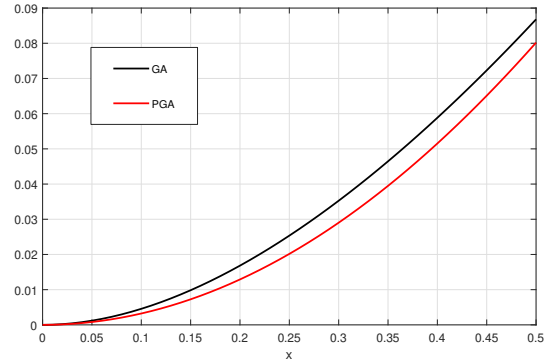}
\vspace{-1 em}
\caption{Comparison of eq. \eqref{eq:11} for GA and PGA with x $\in$ (0,..,0.5)}
\end{center}
\label{fig:funcao_phi_multi6}
\end{figure}

This difference observed between these functions provides a performance improvement that we will see in the next section.
In Algorithm 1 we have the description of the proposed PGA construction algorithm with $\phi_p$ in \eqref{eq:17}. For the calculation of the function $\phi_p^{-1}$, inverse function of \eqref{eq:17}, by the bisection method, an approach optimized is presented in Algorithm 2.

\begin{algorithm}
\caption{PGA construction}
\SetAlgoLined
\KwIn{$N$, code length}
\KwIn{$K$, information bits}
\KwIn{design-SNR $E_{dB}=(RE_b/N_o)$ in dB}
\KwOut{$F \in \{0,1,\ldots,N-1\}$ with
$|F|=N$ }
 $S=10^{EdB/10}$\;
 $n=log_2N$\;
 $W \in R^N$$, W(0)=4S$\;
 \For{i = 1 to $n$}{
  $d=2^i$ \;
  \For{j = 1 to $\frac{d}{2}-1$}{
   $W(j) = \phi_p^{-1}(1-(1-\phi_p(W(j-1)))^2)$\;
   $W(d/2+j) = 2W(j-1)$\;
  }
 }
$F={\rm Sorts ~W ~indices ~in ~ascending ~order}$\;
\end{algorithm}

\begin{algorithm}
\caption{$\phi_p^{-1} function$}
\SetAlgoLined
\KwIn{$y$, value input}
\KwOut{$x$, value output}
  \If{y = 0}{
     x = 0\;
     return\;
     }
  aux = 1\;
  base = $\phi_p$(aux)\;
  iteration-times = 20\;
  \eIf{y $\leq$ base}{
     \While{y $<$ base}{
         aux = aux*0.1\;
         base = $\phi_p$(aux)\;
         }
     anchor1 = aux\;
     anchor2 = aux*10\;
     \For{i = 1 to iteration-times}{
         x = (anchor1+anchor2)/2\;
         aux = $\phi_p$(x)\;
         \eIf{y $<=$ aux}{
            anchor2 = x\;
         }{
            anchor1 = x\;
        }
     }
  }{
     \While{y $>=$ base}{
         aux = aux + 10\;
         base = $\phi_p$(aux)\;
     }
     anchor1 = aux - 10\;
     anchor2 = aux\;
     \For{i = 1 to iteration-times}{
         x = (anchor1+anchor2)/2\;
         aux = $\phi_p$(x)\;
         \eIf{y $<=$ aux}{
            anchor2 = x\;
         }{
            anchor1 = x\;
         }
     }
}
\end{algorithm}

\section{Simulations}

In this section, we evaluate the proposed PGA construction algorithm and compare it against existing approaches for several scenarios. Initially,  we present in table 1 the difference in channels between the GA construction and the proposed PGA construction, for $R \in (1/2,1/3,2/3)$ and design-SNR of 1dB and AWGN channel. This difference in channels is due to the adjustment provided by the piecewise method applied directly to the polar code construction.

\begin{table}[htb]
\centering
\caption{Channel difference between PGA and GA}
\begin{tabular}{|c|c|c|c|c|c|}
\hline
R & 128  & 256  & 512  & 1024 & 2048 \\ \hline\hline
1/2 & 4    & 2    & 2    & 8    & 18   \\ \hline
1/3 & 0    & 2    & 4    & 10   & 18   \\ \hline
2/3 & 0    & 2    & 4    & 10   & 28   \\ \hline
\end{tabular}
\end{table}

We also observed in table 1 that the difference in channels tends to be greater with the increase in the codeblock length $N$. In the following, we illustrate the results of Monte Carlo simulations using the AFF3CT toolbox as a library \cite{Cassagne}. The simulations employ Binary Phase shift keying (BPSK) modulation and the AWGN channel, the SC decoder and code designs with design-SNR of 1dB. The simulation loops had as stop criterion the target Frame Error Rate (FER), FER$ = 200$.





In Fig. 6, we have compared the proposed PGA with GA for $R = 1/2$ and several codeblock lengths. We can notice that for $N \geq 128$ the PGA has an increasing gain in FER for all codeblock lengths, which can reach $0.25$ dB for $N = 2048$.

\begin{figure}[htb]
\vspace{-0.45em}
\begin{center}
\includegraphics[scale=0.45]{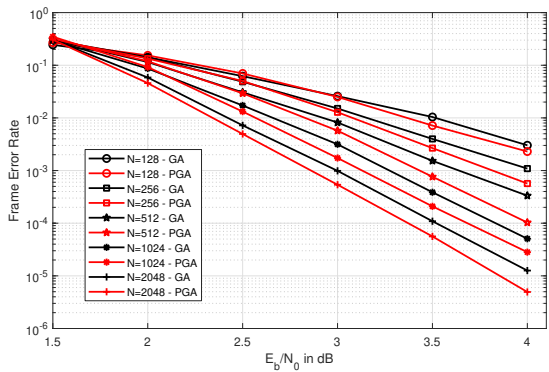}
\vspace{-1 em}
\caption{FER performance of GA and PGA for $R=1/2$.}
\end{center}
\label{fig:fig_varios2h}
\end{figure}

In Fig. 7, we compare the proposed PGA with GA for $R=1/3$ and various codeblock lengths. We observe that the PGA has a FER gain for codeblock lengths $N \geq 256$, reaching $0.15$dB for $N=2048$ . In the block length $N=128$ there is no channel difference between the PGA and GA constructions, as can be seen in table 1.

\begin{figure}[htb]
\vspace{-0.5em}
\begin{center}
\includegraphics[scale=0.45]{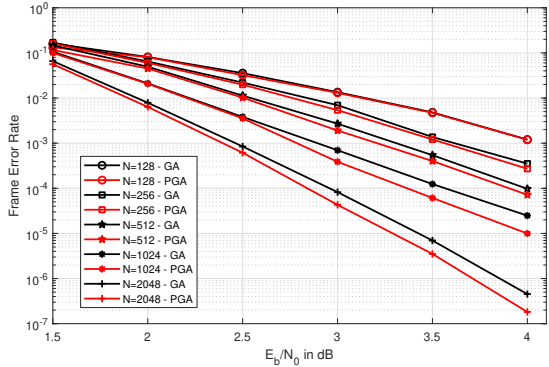}
\vspace{-1 em}
\caption{Comparative FER performance between GA and PGA, for $R=1/3$.}
\end{center}
\label{fig:fig_variosK13e}
\end{figure}

In Fig. 8, the same comparison between PGA and GA is made, now for $R=2/3$ and several block lengths. In this scenario, we verify that PGA has a FER gain from $N \geq 512$, reaching $0.25$dB for $N=2048$. At codeblock length $N=128$ there is no channel difference between PGA and GA constructions, as can be seen in table 1. For codeblock length $N=256$, even with a difference of 2 channels according to table 1, the performance is the same between PGA and GA constructions.


\begin{figure}[htb]
\vspace{-0.5em}
\begin{center}
\includegraphics[scale=0.45]{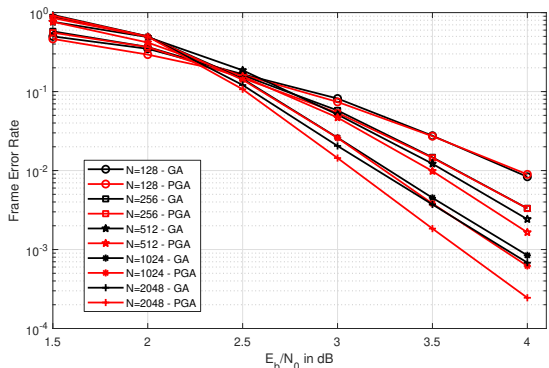}
\vspace{-1 em}
\caption{Comparative FER performance between GA and PGA, for $R=2/3$.}
\end{center}
\label{fig:fig_variosK23e}
\end{figure}

In general, as the codeblock length $N$ increases, we observe a continuous improvement in the performance FER provided by PGA when compared to GA, and for short codeblock length $N$, in cases where there is a difference in channels, there is no difference in FER performance.
For $N=1024$ and $N=2048$, $R=1/2$; in Fig. 9 we compare the proposed PGA, GA and the methods for long blocks proposed by Fang \cite{Fang}, Dai \cite{Dai} and Ochiai \cite{Ochiai}. The results show that the methods for long blocks proposed by Fang \cite{Fang}, Dai \cite{Dai} and Ochiai \cite{Ochiai} have the same FER performance as GA.

\begin{figure}[htb]
\vspace{-0.5em}
\begin{center}
\includegraphics[scale=0.45]{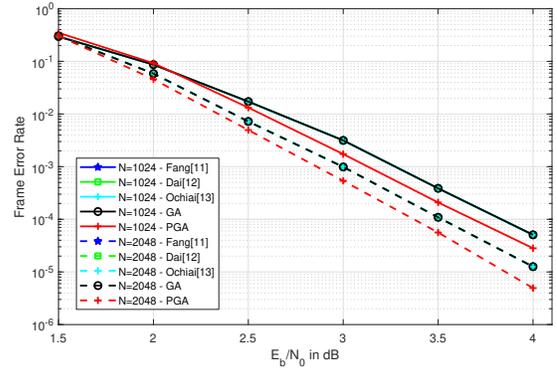}
\vspace{-1 em}
\caption{FER performance between GA, PGA, Fang \cite{Fang}, Dai \cite{Dai} and Ochiai \cite{Ochiai} for $R=1/2$.}
\end{center}
\label{fig:fig_varios_PGA}
\end{figure}

\vspace{-0.5 em}
\section{Conclusions}
\vspace{-0.5 em}
We have presented the construction of polar codes based on the PGA approach. In particular, with the piecewise approach we obtain a function that replaces the original GA function with a more accurate approximation, which results in significant gain in performance. The proposed PGA design of polar codes is presented in its integral form as well as an alternative approximation that does not rely on the integral form. The results show that the proposed PGA design outperforms the standard GA for several examples of polar codes and rates.


\vspace{-0.5 em}
\begin{thebibliography}{00}
\vspace{-0.5 em}
\bibitem{Arikan}
E. Arikan, ``Channel Polarization: A Method for Constructing Capacity-Achieving Codes for Symmetric Binary-Input Memoryless Channels", \textit{IEEE Transactions on Information Theory}, vol. 55 no. 7, pp. 3051--3073, July 2009.

\bibitem{5G}
(2018). ``Technical Specification Group Radio Access Network" [Online]. Available: \textit{http://www.3gpp.org/ftp/Specs/archive/38series/38.212/}

\bibitem{Mori1}
R. Mori and T. Tanaka, ``Performance of polar codes with the construction using density evolution", \textit{IEEE Communications Letters}, vol. 13, no. 7, pp. 519--521, July 2009.

\bibitem{Mori2}
R. Mori and T. Tanaka, ``Performance and construction of polar codes on symmetric binary-input memoryless channels", \textit{International Symposium on Information Theory (ISIT)}, 2009, pp. 1496--1500.

\bibitem{Tal1}
I. Tal and A. Vardy, ``How to Construct polar codes", \textit{IEEE Transactions on Information Theory}, vol. 59, no. 10, pp. 6562--6582, 2013.

\bibitem{Chung}
Sae-Young Chung, T. J. Richardson and R. L. Urbanke, ``Analysis of sum-product decoding of low-density parity-check codes using a Gaussian approximation,” \textit{IEEE Transactions on Information Theory}, vol. 47, no. 2, pp. 657--670, Feb 2001.

\bibitem{Trifonov}
P. Trifonov, ``Efficient design and decoding of polar codes,” \textit{IEEE Transactions on Communications}, vol. 60, no. 11, pp. 1--7, 2012.

\bibitem{RZhang}
G. He, J. Belfiore, I. Land, G. Yang, X. Liu, Y. Chen, R. Li, J. Wang, Y. Ge, R. Zhang, W. Tong, ``B-expansion: A Theoretical Framework for Fast and Recursive Construction of Polar Codes", \textit{IEEE Global Communication Conference (GLOBECOM 2017)}, pp. 1--6, 2017.

\bibitem{Schurch}
C. Schurch, ``A partial Order for the Synthesized Channels of a Polar Code", \textit{IEEE International Symposium on Information Theory (ISIT)}, pp. 220--224, 2016.

\bibitem{Jeongseok}
Jeongseok Ha, Jaehong Kim and S. W. McLaughlin, ``Rate-compatible puncturing of low-density parity-check codes,” \textit{IEEE Transactions on Information Theory}, vol. 50, no. 11, pp. 2824--2836, Nov. 2004.

\bibitem{Fang}
Z. Fang, J. Gao and R. Liu, ``A simplified Gaussian approximation algorithm for polar codes,” \textit{3rd IEEE International Conference on Computer and Communications (ICCC)}, pp. 2429--2433, 2017.

\bibitem{Dai}
J. Dai, K. Niu, Z. Si, C. Dong and J. Lin, ``Does Gaussian Approximation Work Well for the Long-Length Polar Code Construction?,” \textit{IEEE Access}, vol. 5, pp. 7950--7963, 2017.

\bibitem{Ochiai}
H. Ochiai, P. Mitran and H. Vincent Poor, ``Capacity-Approaching Polar Codes with Long Codewords and Successive Cancellation Decoding Based on Improved Gaussian Approximation,” \textit{IEEE Transactions on Communications}, vol. 69, no. 1, pp. 31--43, 2021.

\bibitem{Vangala}
H. Vangala, E. Viterbo and Y. Hong, ``A Comparative Study of Polar Code Constructions for the AWGN Channel", \textit{https://arxiv.org/pdf/1501.02473.pdf}, Jan 2015.

\bibitem{Cheng}
J. Li, M. Hu and Z. Cheng, ``Research on Polar Code Construction Algorithms under Gaussian Channel", \textit{2018 Tenth International Conference on Ubiquitous and Future Networks (ICUFN)}, pp. 515--518, July 2018.

\bibitem{Oliveira1}
R. M. Oliveira and R. C. de Lamare, ``Puncturing Based on Polarization for Polar Codes in 5G Networks", \textit{2018 15th International Symposium on Wireless Communication Systems (ISWCS)}, pp. 1--5, 2018.

\bibitem{Oliveira2}
R. M. Oliveira and R. C. de Lamare, ``Rate-Compatible Polar Codes Based on Polarization-Driven Shortening", \textit{IEEE Communications Letters}, vol. 22, no. 10, pp. 1984--1987, 2018.

\bibitem{Oliveira3}
R. M. Oliveira and R. C. de Lamare, ``Non-Uniform Channel Polarization and Design of Rate-Compatible Polar Codes", \textit{16th International Symposium on Wireless Communication Systems (ISWCS)}, Oulu, Finland, pp. 537-541, 2019.

\bibitem{Oliveira4}
R. M. Oliveira and R. C. De Lamare, ``Design of Rate-Compatible Polar Codes Based on Non-Uniform Channel Polarization", \textit{IEEE Access}, vol. 9, pp. 41902--41912, 2021.

\bibitem{Cassagne}
A. Cassagne, O. Hartmann, M. Leonardon, K. He, C. Leroux, R. Tajan, O. Aumage, D. Barthou, T. Tonnellier, V. Pignoly, B. Le Gal, and C. Jego, ``Aff3ct: A fast forward error correction toolbox!,” \textit{Elsevier SoftwareX}, vol. 10, p. 100345, Oct. 2019. [Online]. Available: http://www.sciencedirect.com/science/article/pii/S2352711019300457

\bibitem{bfpeg}
A. G. D. Uchoa, C. Healy, R. C. de Lamare and R. D. Souza, "Design of LDPC Codes Based on Progressive Edge Growth Techniques for Block Fading Channels," in IEEE Communications Letters, vol. 15, no. 11, pp. 1221-1223, November 2011, doi: 10.1109/LCOMM.2011.092911.111520.

\bibitem{rrser}
Y. Cai, R. C. de Lamare, B. Champagne, B. Qin and M. Zhao, "Adaptive Reduced-Rank Receive Processing Based on Minimum Symbol-Error-Rate Criterion for Large-Scale Multiple-Antenna Systems," in IEEE Transactions on Communications, vol. 63, no. 11, pp. 4185-4201, Nov. 2015.

\bibitem{rootldpc}
A. G. Uchoa, C. T. Healy, R. C. de Lamare, Structured root-LDPC codes and PEG-based techniques for block-fading channels. J Wireless Com Network 2015, 213 (2015).

\bibitem{memd}
C. T. Healy and R. C. de Lamare, "Design of LDPC Codes Based on Multipath EMD Strategies for Progressive Edge Growth," in IEEE Transactions on Communications, vol. 64, no. 8, pp. 3208-3219, Aug. 2016.

\bibitem{baplnc}
J. Gu, R. C. de Lamare and M. Huemer, "Buffer-Aided Physical-Layer Network Coding With Optimal Linear Code Designs for Cooperative Networks," in IEEE Transactions on Communications, vol. 66, no. 6, pp. 2560-2575, June 2018.

\bibitem{dopeg}
C. T. Healy and R. C. de Lamare, "Decoder-Optimised Progressive Edge Growth Algorithms for the Design of LDPC Codes with Low Error Floors," in IEEE Communications Letters, vol. 16, no. 6, pp. 889-892, June 2012.

\bibitem{jidf}
R. C. de Lamare and R. Sampaio-Neto, "Adaptive Reduced-Rank Processing Based on Joint and Iterative Interpolation, Decimation, and Filtering," in IEEE Transactions on Signal Processing, vol. 57, no. 7, pp. 2503-2514, July 2009.

\bibitem{spa}
R. C. De Lamare and R. Sampaio-Neto, "Minimum Mean-Squared Error Iterative Successive Parallel Arbitrated Decision Feedback Detectors for DS-CDMA Systems," in IEEE Transactions on Communications, vol. 56, no. 5, pp. 778-789, May 2008.

\bibitem{mfsic}
P. Li, R. C. de Lamare and R. Fa, "Multiple Feedback Successive Interference Cancellation Detection for Multiuser MIMO Systems," in IEEE Transactions on Wireless Communications, vol. 10, no. 8, pp. 2434-2439, August 2011.

\bibitem{dfcc}
P. Li and R. C. De Lamare, "Adaptive Decision-Feedback Detection With Constellation Constraints for MIMO Systems," in IEEE Transactions on Vehicular Technology, vol. 61, no. 2, pp. 853-859, Feb. 2012.

\bibitem{mbdf}
R. C. de Lamare, "Adaptive and Iterative Multi-Branch MMSE Decision Feedback Detection Algorithms for Multi-Antenna Systems," in IEEE Transactions on Wireless Communications, vol. 12, no. 10, pp. 5294-5308, October 2013.

\bibitem{mmimo}
R. C. de Lamare, "Massive MIMO systems: Signal processing challenges and future trends," in URSI Radio Science Bulletin, vol. 2013, no. 347, pp. 8-20, Dec. 2013.

\bibitem{mbthp}
K. Zu, R. C. de Lamare and M. Haardt, "Multi-Branch Tomlinson-Harashima Precoding Design for MU-MIMO Systems: Theory and Algorithms," in IEEE Transactions on Communications, vol. 62, no. 3, pp. 939-951, March 2014.

\bibitem{wence}
W. Zhang et al., "Large-Scale Antenna Systems With UL/DL Hardware Mismatch: Achievable Rates Analysis and Calibration," in IEEE Transactions on Communications, vol. 63, no. 4, pp. 1216-1229, April 2015.


\bibitem{bfidd}
A. G. D. Uchoa, C. T. Healy and R. C. de Lamare, "Iterative Detection and Decoding Algorithms for MIMO Systems in Block-Fading Channels Using LDPC Codes," in IEEE Transactions on Vehicular Technology, vol. 65, no. 4, pp. 2735-2741, April 2016.

\bibitem{vfap}
J. Liu and R. C. de Lamare, "Low-Latency Reweighted Belief Propagation Decoding for LDPC Codes," in IEEE Communications Letters, vol. 16, no. 10, pp. 1660-1663, October 2012.

\bibitem{kaids}
C. T. Healy, Z. Shao, R. M. Oliveira, R. C. de Lamare, L. L. Mendes,  'Knowledge-aided informed dynamic scheduling for LDPC decoding of short blocks', IET Communications, 2018, 12, (9), p. 1094-1101

\bibitem{1bitidd}
Z. Shao, R. C. de Lamare and L. T. N. Landau, "Iterative Detection and Decoding for Large-Scale Multiple-Antenna Systems With 1-Bit ADCs," in IEEE Wireless Communications Letters, vol. 7, no. 3, pp. 476-479, June 2018.

\bibitem{did}
P. Li and R. C. de Lamare, "Distributed Iterative Detection With Reduced Message Passing for Networked MIMO Cellular Systems," in IEEE Transactions on Vehicular Technology, vol. 63, no. 6, pp. 2947-2954, July 2014.

\bibitem{lrcc}
H. Ruan and R. C. de Lamare, "Distributed Robust Beamforming Based on Low-Rank and Cross-Correlation Techniques: Design and Analysis," in IEEE Transactions on Signal Processing, vol. 67, no. 24, pp. 6411-6423, 15 Dec.15, 2019.

\bibitem{aaidd}
R. B. Di Renna and R. C. de Lamare, "Adaptive Activity-Aware Iterative Detection for Massive Machine-Type Communications," in IEEE Wireless Communications Letters, vol. 8, no. 6, pp. 1631-1634, Dec. 2019

\bibitem{listmtc}
R. B. Di Renna and R. C. de Lamare, "Iterative List Detection and Decoding for Massive Machine-Type Communications," in IEEE Transactions on Communications, vol. 68, no. 10, pp. 6276-6288, Oct. 2020.

\bibitem{dynovs}
Z. Shao, L. T. N. Landau and R. C. de Lamare, "Dynamic Oversampling for 1-Bit ADCs in Large-Scale Multiple-Antenna Systems," in IEEE Transactions on Communications, vol. 69, no. 5, pp. 3423-3435, May 2021.

\bibitem{rcpd}
R. M. Oliveira and R. C. de Lamare, "Rate-Compatible Polar Codes Based on Polarization-Driven Shortening," in IEEE Communications Letters, vol. 22, no. 10, pp. 1984-1987, Oct. 2018.

\bibitem{detmtc}
R. B. Di Renna, C. Bockelmann, R. C. de Lamare and A. Dekorsy, "Detection Techniques for Massive Machine-Type Communications: Challenges and Solutions," in IEEE Access, vol. 8, pp. 180928-180954, 2020.

\bibitem{srbars}
X. Lu and R. C. de Lamare, "Opportunistic Relaying and Jamming Based on Secrecy-Rate Maximization for Multiuser Buffer-Aided Relay Systems," in IEEE Transactions on Vehicular Technology, vol. 69, no. 12, pp. 15269-15283, Dec. 2020.

\bibitem{dynmtc}
R. B. Di Renna and R. C. de Lamare, "Dynamic Message Scheduling Based on Activity-Aware Residual Belief Propagation for Asynchronous mMTC," in IEEE Wireless Communications Letters, vol. 10, no. 6, pp. 1290-1294, June 2021.

\bibitem{nupd}
R. M. Oliveira and R. C. De Lamare, "Design of Rate-Compatible Polar Codes Based on Non-Uniform Channel Polarization," in IEEE Access, vol. 9, pp. 41902-41912, 2021.



\end{thebibliography}
\end{document}